\documentclass[sigconf]{acmart}

\usepackage{graphicx}
\usepackage{algorithm}
\usepackage{algorithmicx}  
\usepackage{algpseudocode}
\usepackage{xcolor}
\usepackage[caption=false]{subfig}
\usepackage{balance}

\graphicspath{{figures/}}

\floatname{algorithm}{Algorithm}

\algrenewcommand\algorithmicforall{\textbf{for each}}

\AtBeginDocument{%
  \providecommand\BibTeX{{%
    \normalfont B\kern-0.5em{\scshape i\kern-0.25em b}\kern-0.8em\TeX}}}

\copyrightyear{2020} 
\acmYear{2020} 
\setcopyright{acmcopyright}
\acmConference[CIKM '20]{Proceedings of the 29th ACM International Conference on Information and Knowledge Management}{October 19--23, 2020}{Virtual Event, Ireland}
\acmBooktitle{Proceedings of the 29th ACM International Conference on Information and Knowledge Management (CIKM '20), October 19--23, 2020, Virtual Event, Ireland}
\acmPrice{15.00}
\acmDOI{10.1145/3340531.3411924}

\begin{document}
\fancyhead{}
\title{G-CREWE: Graph CompREssion With Embedding for Network Alignment}

\author{Kyle K. Qin}
\affiliation{%
  \institution{RMIT University}
}
\email{kai.qin2@rmit.edu.au}

\author{Flora D. Salim}
\affiliation{%
  \institution{RMIT University}
}
\email{flora.salim@rmit.edu.au}

\author{Yongli Ren}
\affiliation{%
  \institution{RMIT University}
}
\email{yongli.ren@rmit.edu.au}

\author{Wei Shao}
\affiliation{%
  \institution{RMIT University}
}
\email{wei.shao@rmit.edu.au}

\author{Mark Heimann}
\affiliation{%
  \institution{University of Michigan}
}
\email{mheimann@umich.edu}

\author{Danai Koutra}
\affiliation{%
  \institution{University of Michigan}
}
\email{dkoutra@umich.edu}


\begin{abstract}
Network alignment is useful for multiple applications that require increasingly large graphs to be processed. Existing research approaches this as an optimization problem or computes the similarity based on node representations. However, the process of aligning every pair of nodes between relatively large networks is time-consuming and resource-intensive. In this paper, we propose a framework, called G-CREWE (\textbf{G}raph \textbf{C}omp\textbf{RE}ssion \textbf{W}ith \textbf{E}mbedding) to solve the network alignment problem. G-CREWE uses node embeddings to align the networks on two levels of resolution, a fine resolution given by the original network and a coarse resolution given by a compressed version, to achieve an efficient and effective network alignment. The framework first extracts node features and learns the node embedding via a Graph Convolutional Network (GCN). Then, node embedding helps to guide the process of graph compression and finally improve the alignment performance. As part of G-CREWE, we also propose a new compression mechanism called MERGE (\textbf{M}inimum D\textbf{E}g\textbf{R}ee Nei\textbf{G}hbors Compr\textbf{E}ssion) to reduce the size of the input networks while preserving the consistency in their topological structure. Experiments on all real networks show that our method is more than twice as fast as the most competitive existing methods while maintaining high accuracy. 
\end{abstract}




\maketitle

\section{Introduction}
A variety of information has been encapsulated with graphs and networks which can naturally represent the roles of entities and their relationships, such as users communication in social networks \cite{chierichetti2009compressing}, mobility of transport or human \cite{shao2017traveling,qin2019solving}, and protein$-$protein interaction \cite{bayati2013message}. Numerous researches have hitherto been established for mining knowledge on graphs, two important problems are network alignment and graph compression. 

Network alignment focuses on inferring the node correspondences between different graphs based on their topological or feature similarity. For instance, it matches users between different social platforms for items recommendation across networks \cite{zhang2016final}. In image recognition, graph matching is used to identify similar objects in different images that are typically denoted by a set of vertices and edges \cite{guo2018neural}. On the other hand, graph compression has become a necessity in the era of big data. Large graphs and networks are converted into smaller ones while preserving their certain characteristics, which could effectively diminish the time and space complexity in different data mining tasks. Hence, one question that we consider is how to apply compression to achieve a more effective and efficient network alignment.

In recent years, many state-of-the-art approaches have been developed to learn node representations or embedding based on graph structure for node classification, link prediction or graph alignment \cite{perozzi2014deepwalk,grover2016node2vec,ribeiro2017struc2vec,heimann2018regal}. Moreover, some studies intend to extract graph embedding from the relation in knowledge graph \cite{wang2020relation} or connections in heterogeneous graph \cite{ren2017location}. Graph Convolutional Networks (GCN) generalize neural networks to work on arbitrarily structured graphs and are a powerful approach in learning node representations \cite{chen2018harp,kipf2016semi}. Using GCNs for network alignment holds promise, but another challenge is that generally network alignment is a resource-intensive operation. With graph compression, networks can be condensed into relative small ones while maintaining the consistency of their topological structure, which could boost learning tasks. Therefore, we show how to apply compression to achieve a more effective and efficient network alignment. 

Our proposed framework is called G-CREWE. It effectively and efficiently generates node embedding using a GCN model and applies a new graph compression technique to accelerate the overall speed of alignment process. However, there are several challenges to deal with when we consider using graph compression. First, a unsuitable graph compression may consume a big amount of computational time itself, we need to ensure the total runtime to perform network alignment is indeed reduced. Second, most graph compression algorithms focus on preserving the weight or path distance between nodes within a single graph, but the ability of maintaining the topological consistency between different networks is rarely studied. Third, we must explore an approach which could effectively match nodes in two compressed networks and the merged nodes between supernodes. The main contributions of this paper are as follows:
\begin{itemize}
\item \textbf{Problem Definition.} We define a problem for enhancing network alignment via both graph compression and embedding, with the aim to achieve a more rapid alignment for the nodes in different coarsened networks while retaining topological consistency during the compression. After the matching between supernodes, the merged nodes are further examined for complete node alignments.
\item \textbf{Algorithms} We propose G-CREWE, an effective framework which integrates the alignment and the compression process by leveraging the power of nodes embedding. A key part of G-CREWE is MERGE, a new compression mechanism that preserves the topological structure of the original graph. Our approach can be applied to attributed and unattributed graphs with virtually no change in formulation, and is unsupervised: it does not require prior alignment information to find high-quality matching.
\item \textbf{Evaluations.} We evaluate the proposed algorithm by intensive experiments to show its efficiency: how it satisfies a balance of important properties, namely accuracy and fast runtime. We analyze how the compression algorithm can keep the topological consistency in disjoint networks, and maintain the efficiency of the alignment process. 
\end{itemize}

The rest of the paper is organized as follows. Section \ref{sec:problem_def} defines the problem of network alignment with compression. Section \ref{sec:framework} presents the proposed solution and its components. Section \ref{sec:experiment} shows the experimental results. Finally, related work and conclusion are given in Section \ref{sec:relate_work} and Section \ref{sec:conclusion}, respectively. The source code of G-CREWE is available at https://github.com/cruiseresearchgroup/G-CREWE.

\section{Problem Definitions}
\label{sec:problem_def}
The main symbols and notations used throughout the paper are summarized in Table \ref{symbols}. $G_i$ and $G'_i$ denote an input network and its compressed version, respectively. A network $G_i(V_i, E_i, F_i)$ has a set of nodes $V_i$, a set of the edges $E_i$ and the attributes $F_i$ for the nodes. $G'_i(V'_i, E'_i, F'_i)$ is the compressed version of $G_i$. Particularly, $G'_i$ contains a set of uncompressed original nodes $C'_i$ and a set of supernodes $U'_i$. Therefore, $V'_i = \{C'_i \cup U'_i\}$. 

\begin{table}[tb]
    \caption{Symbols and Notations}
    \begin{tabular}{ll}
    \toprule
    \textbf{Symbols} & \textbf{Definitions} \\
    \hline
    $G_i, G'_i$ & original graph and its compressed version \\
    $V_i, V'_i$ & set of the nodes in $G_i$ and $G'_i$ \\
    $E_i, E'_i$ & set of the edges in $G_i$ and $G'_i$ \\
    $F_i, F'_i$ & attributes of the nodes in $G_i$ and $G'_i$ \\
    $C'_i$ & set of uncompressed original nodes in $G'_i$ \\
    $U'_i$ & set of the supernodes in $G'_i$ \\
    $|V_i|, |V'_i|, |C'_i|, |U'_i|$ & \# of the nodes in sets $V_i$, $V'_i$, $C'_i$ and $U'_i$  \\
    \hline 
    $A_i, A'_i$ & adjacency matrix of $G_i$ and $G'_i$ \\
    $N(v)$ & neighbors of node $v$ \\
    $ \delta(N(v))$ & minimum degree in the neighbors of $v$ \\
    $\Delta(G_i), \delta(G_i)$ & maximum or minimum degree of $G_i$ \\
    $\varphi$ & compression ratio $\varphi = 1- \frac{|V'_i|}{|V_i|}$ \\
    $K$ & maximum hop distance considered \\
    $\gamma$ & hop discount factor \\
    $\eta$ & node degree threshold \\
    $\lambda$ & number of top nodes to fast pairing \\
    $\omega$ & nodes similarity threshold \\
    $p$ & dimension of node structural embedding \\
    $\gamma_{1}$, $\gamma_{2}$ & weights of structural and attribute features \\ 
    $\alpha$ & number of top alignments for each node \\
    \bottomrule 
    \end{tabular}
    \label{symbols}
\end{table}

\begin{figure}[tb]
    \centering
    \includegraphics[width=0.95\linewidth]{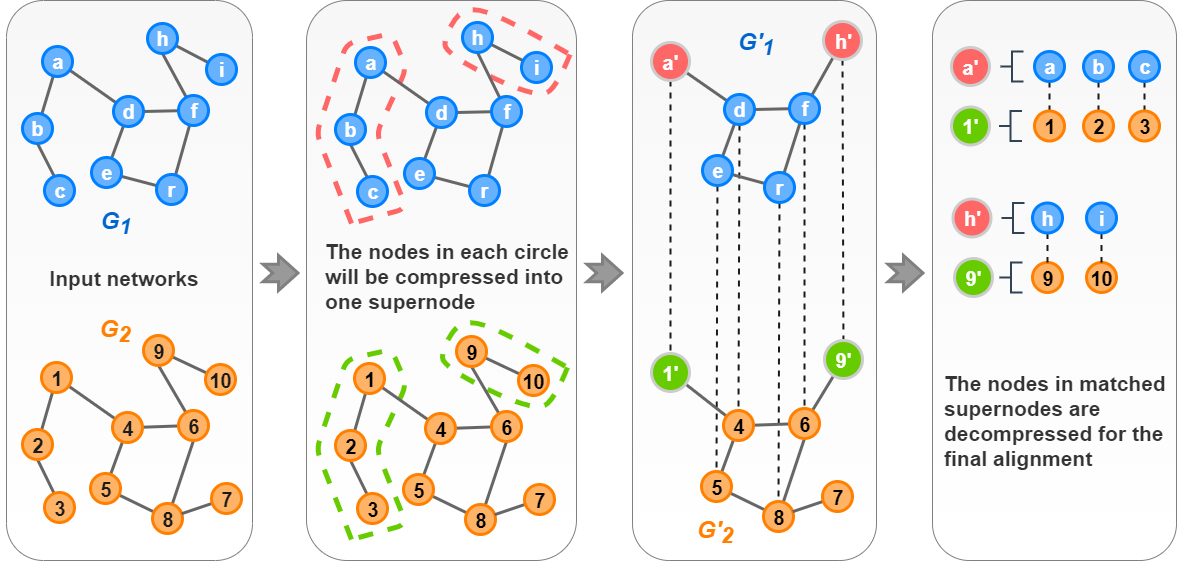}
    \caption{An illustrative example of applying compression on two networks for node alignments.}
    \label{fig:graph_com_align}
\end{figure}

Fig.~\ref{fig:graph_com_align} demonstrates an example of the problem. We have two input networks $G_1$ and $G_2$ for alignment. Graph compression is carried on these networks separately. During the compression, the nodes $a, b, c$ in $G_1$ are merged into the supernode $a'$ and nodes $h, i$ are replaced by a supernode $h'$. Meanwhile, the nodes $1, 2, 3$ in $G_2$ are represented by the supernode $1'$ and both nodes $9, 10$ are replaced by node $9'$. Next, the alignments are conducted between the compressed networks $G'_1$ and $G'_2$, a good result shall match the supernodes $a'$ and $h'$ in $G'_1$ with $1'$ and $9'$ in $G'_2$, respectively. Moreover, the uncompressed nodes $d, e, f, r$ are aligned to nodes $4, 5, 6, 8$ respectively. In the final step, we further align the sub-nodes in the supernodes that have been matched already. As we can see, compression reduces the size of the networks and avoids alignments among entire nodes. The problem can be defined as follows: \\ 
\textbf{Problem 1.} Network alignment with compression and embedding. \\
\textit{\textbf{Given:} (1) two networks $G_1(V_1, E_1)$ and $G_2(V_2, E_2)$ with node-sets $V_1$ and $V_2$, and possibly node attributes $F_1$ and $F_2$ respectively; (2) a compression ratio $\varphi$, 0 $<$ $\varphi$ $<$ 1.} \\
\textit{\textbf{Intermediate}: (1) the embedding of nodes in two original networks; (2) two compressed networks $G'_1$ and $G'_2$ that keep high topologically structural consistency between each other. $G'_1$ and $G'_2$ contain sets of uncompressed original nodes $C'_1$ and $C'_2$, and sets of supernodes $U'_1$ and $U'_2$ respectively.} \\
\textit{\textbf{Output}: (1) $C'_1 \Leftrightarrow C'_2$, the alignments for the uncompressed original nodes between $G'_1$ and $G'_2$; (2) $U'_1 \Leftrightarrow U'_2$, the alignments for the supernodes between two compressed networks; (3) the further match for the sub-nodes in those supernodes that have been highly aligned.}

\section{G-CREWE: The Proposed Framework}
\label{sec:framework}
\begin{figure*}[tb]
    \centering
    \includegraphics[width=0.85\textwidth]{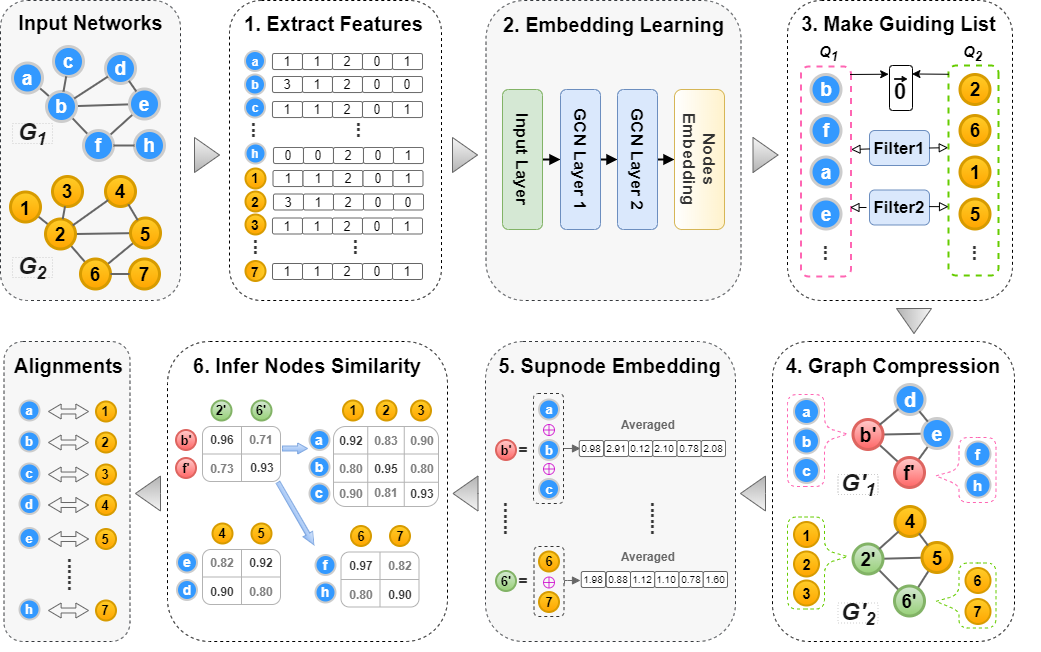}
    \caption{The overview of the proposed framework. (1) Extracting the structural features of nodes in two networks with considering 2-hop neighbors. (2) 2-layer GCN is applied on these features to produce node embedding. (3) Creating guiding-lists $Q_1$ and $ Q_2$ to guide the process of compression. (4) Selecting one node from $Q_1$ or $Q_2$ in order as the starting point for every compression (equals to generate a supernode) in individual network, it repeats until the compression ratio $\varphi$ is met. (5) Calculating the embedding for supernodes via element-wise addition on their sub-nodes with averaging. (6) Inferring the similarity scores between nodes.}
    \label{fig:main_framwork}
\end{figure*}

Fig.~\ref{fig:main_framwork} shows the framework that involves several main processes, including node features extraction, node embedding learning, guiding-list generation, graph compression and inferring node similarity. 

\subsection{Node Features Extraction}
\label{sec:node_fea_extract}
In the first part of the solution, it is crucial to establish some characteristic features of nodes before embedding learning. Inspired by REGAL~\cite{heimann2018regal}, we extract the node identities based on local structure and possible node attributes in the following ways. \\
\textbf{Structural Features.} We use $N^k(v)$ to denote the neighbors of node $v \in V_i$ at $k$ hops away from $v$ in graph $G_i$. We create a vector $d_v^k$ to capture the degree distribution within the $k$-hop neighborhood of $v$.
We bin the node degrees logarithmically in $d_v^k$ such that the $i$-th entry of $d_v^k$ records the number of node $u \in N^k(v)$ where $\lceil log_2(deg(u)) \rceil = i$.  Logarithmic binning is a robust way to capture node degree distributions that also reflects the power-law degree distribution of many real-world graphs~\cite{heimann2018regal}. 
The final structural feature vector $d_v$ for node $v$ is the aggregation of its neighborhoods' features over hop distances up to $K$, weighted by a discount factor $\gamma$: $d_v = \sum_{k=1}^{K}\gamma ^{k-1}d_v^k$; $\gamma \in$ (0, 1]; $K$ is the maximum hop distance; $\gamma$ controls the impact of neighbors at different hops. \\
\textbf{Attribute Features.} When there are $F$ node attributes in input graphs, we first create a $F$ dimensional vector $f_v$ for each node $v$ to store its attributes. The entry $i$ of $f_v$ contains $i^{th}$ attribute value of node $v$. Here, REGAL~\cite{heimann2018regal} keeps the original values of attributes, but we employ one-hot encoding to represent it in $f_v$ if the attribute is categorical. Otherwise, normalisation can be applied on numeric values of attributes that will range between [0, 1].

The $m$-dimensional feature descriptor $x_{v}$ is given by the structural features $d_v$, as all the structural features for each node $v$ among the $n$ nodes of both graphs $G_{1}$ and $G_{2}$ are collected in a matrix $X \in \mathbb{R}^{n \times m}$. Here, $m$ is $log_{2}\Delta(G)$ and $\Delta(G)$ is the maximum degree in input graphs. In addition, the matrix $Y \in \mathbb{R}^{n \times r}$ is formed by attribute features of all $n$ nodes, and $r$ is the final dimension of each node's attribute feature. While xNetMF~\cite{heimann2018regal} embeds nodes by implicitly factorizing a similarity matrix based on these structural and attribute-based node features, we instead learn node embedding by propagating the structural features through local neighborhoods using graph convolutional networks, and then directly combine the node embedding with encoded attribute features in a new way for final similarity computation, as we describe next.   

\subsection{Structural Embedding Learning}
\label{sec:node_emb_learn}
Graph Convolutional Networks are the specialized neural models that can operate directly on arbitrarily structured graphs for semi-supervised classification or other learning tasks. A well-established GCN is capable of learning representations through multiple layers to encode both local graph structure and node features in an irregular network, such as social network or protein-interaction network \cite{kipf2016semi}. To solve our problem, we learn node embedding across two networks simultaneously by a GCN model which then is used for graph compression and alignment problem.

We got the feature matrix $X$ that holds significant features for each node in both graphs $G_{1}$ and $G_{2}$ in Section \ref{sec:node_fea_extract}. Then, a fast multi-layer GCN \cite{kipf2016semi} is adopted to produce embedding for these nodes with $X$ as input. The layer-wise propagation rule of the model is given below:
\begin{equation}
H^{(l+1)} = \sigma (\widetilde{D}^{-\frac{1}{2}} \widetilde{A}_{join} \widetilde{D}^{-\frac{1}{2}} H^{(l)} W^{(l)})
\end{equation}
Here, $\widetilde{A}_{join} = A_{join} + I$ is the matrix $A_{join} \in \mathbb{R}^{n \times n}$ with adding self-connections and $A_{join}$ is the vertical concatenation of two adjacency matrix of input graphs. This can help to learn nodes embedding of two graphs in a consistent space via a single model. Moreover, $I$ is the identity matrix, $\widetilde{D}$ is the diagonal node degree matrix of $\widetilde{A}_{join}$, $W^{(l)}$ is a trainable weight matrix in the $l^{th}$ neural graph layer, $\sigma$ denotes a non-linear activation function like the ReLU or Tanh, and $H^{(l)}$ is the matrix of activations in the $l^{th}$ layer. Specifically speaking, $H^{(0)}$ is the feature matrix $X$ and the initial $W^{(l)}$ in each layer is generated at random. 

Without any previous training, we employ a multi-layer GCN to yield meaningful embedding for nodes. Previous work had shown that even a GCN model with \emph{random} weights serves as a powerful feature extractor for a graph akin to Weisfeiler-Lehman algorithm~\cite{kipf2016semi}.  Additionally, if some node matchings are known a priori, we can use the loss function between the embedding difference of matching nodes to train the GCN. As our present focus is on \emph{unsupervised} graph alignment, we leave this \emph{semi-supervised} setting for future work.
As an example, a 2-layer model is shown as follows:
\begin{equation}
Z = Tanh (\widehat{A}_{join} Tanh (\widehat{A}_{join} X W^{(0)}) W^{(1)})
\label{equ:2_layers_con}
\end{equation}
Here, $\widehat{A}_{join} = \widetilde{D}^{-\frac{1}{2}} \widetilde{A}_{join} \widetilde{D}^{-\frac{1}{2}}$ is first computed. And weight matrix in each layer is initialized via the approach introduced by Glorot and Bengio \cite{glorot2010understanding}. $W^{(0)} \in \mathbb{R}^{m \times h}$ is an input-to-hidden weight matrix for the first hidden layer and $W^{(1)} \in \mathbb{R}^{h \times p}$ is a hidden-to-output weight matrix. $Z \in \mathbb{R}^{n \times p}$ is the output matrix which contains the representation for the nodes in both graphs. Eventually, $Z$ is split into $Z_1$ and $Z_2$ for the nodes in $V_{1}$ and $V_{2}$ respectively. $p$, the dimension of node embedding, is configured to $log_{2}\Delta(G) \times 2$ in our practice and $\Delta(G)$ is the maximum degree of the input graphs. In this implementation, we can produce the embedding for nodes across two graphs in a shared space without actual training by propagating neighbor features with weights in different layers. 
\\
\textbf{Supernode Embedding.} Note that the GCN is only utilized to learn embedding of nodes from original graphs. As we mentioned before, a certain number of supernodes will be produced in each graph after the process of compression which is presented in the following section. For attaining the embedding of one supernode, we conduct element-wise addition on the embedding vectors of all the sub-nodes merged in it, with averaging the values. Assumed that a supernode A = \{a,b,c\} has three sub-nodes from original graph, its embedding is calculated with the equation as follows:
\begin{equation}
vec_A = \frac{\sum_{i \in A} vec_i}{|A|}
\label{equ:sub_embedding}
\end{equation}

\subsection{Minimum Degree Neighbors Compression}
\label{sec:MERGE}
In this section, we introduce a method called MERGE (\textbf{M}inimum D\textbf{E}g\textbf{R}ee Nei\textbf{G}hbors Compr\textbf{E}ssion) to efficiently reduce graphs with a defined compression ratio $\varphi$. The compressed graph with less nodes and edges could contribute to smaller time complexity and consumption of computational resources. Intuitively, clustering-based graph summarization methods have a chance to assign similar nodes into corresponding groups for compression purpose \cite{riondato2017graph}. In the practice, traditional clustering algorithms like K-mean and DBSCAN \cite{rokach2005clustering} lack the ability to discover each pair of starting points for executing consistent compression on both graphs, with satisfying a defined compression ratio. In addition, several popular clustering methods are particularly time-consuming for assembling a huge number of nodes in a space due to the heavy comparison. On contrast, MERGE is a lightweight method that designed to condense multiple input graphs with several advantages, such as reserving structural consistency among graphs in compression, fast execution and easy control of compression ratio. The steps of the algorithm are demonstrated in Algorithm \ref{alg:Main_alg}.

The compression method accepts one graph as input each time. It is assumed that $G_1(V_1, E_1)$ is one of the original graphs for alignment and $G'_1$ is the compressed version. We first define a compression ratio $\varphi$ to denote the percentage of the nodes in an original graph that need compression or reduction. Also, a guiding-list $Q_1$ is made for supervising the selection of one node as starting point in each compression operation (equals outputting a supernode), where $Q_1$ contains a subset of the nodes $\in$ $V_1$ in certain order and the process of its construction is explained next. For instance, when a node $v_{start}$ is chosen from the top of $Q_1$ for one compression, we retrieve all the neighbors $N(v_{start})$ of $v_{start}$ in $G'_1$ with current compressed state. It means that $N(v_{start})$ can be comprised of either uncompressed original nodes or supernodes. Then, the nodes with minimum degree $\delta(N(v_{start}))$ among $N(v_{start})$ are reserved into list $L$ for the upcoming merging. Next, we find all the neighbors $N(L)$ of each node in $L$ and then compress the nodes in $L$ by removing them from the current $G'_1$ along with the edges incident to them, and adding one new supernode $v'$ with an edge linking to each node $u \in \{N(L) - L\}$. We then update $Q_1$ by deleting away the nodes that have been merged and appending $v'$ to it. In addition, a dictionary $T_1$ is employed to store each supernode $v'$ along with its sub-nodes - the ones in $L$. However, if the node $v \in$ $L$ is also a supernode, we assign its sub-nodes to $v'$ instead of $v$ itself and then remove $v$ from $T_1$. This compression operation is completed and new cycles repeat until the compression ratio $\varphi$ is met. Fig. \ref{fig:main_framwork} illustrates a result about compressing two graphs in the fourth stage of the framework.
\\
\textbf{Guiding-list Generation.} As we mentioned above, the guiding-list $Q_i$ is important to instruct the process of compression on each graph. It enumerates the special nodes that can be used as a starting point for each compression operation. We maintain two storage lists $L_1$ and $L_2$ which initially hold every node $v$ for $V_1$ and $V_2$ respectively, and from which we filter out candidates to add to the guiding-lists $Q_1$ and $Q_2$ for two graphs. 

Our selection process of guiding-list candidates uses three mechanisms, as follows. We notice that the nodes with larger degree tend to be better starting points for compression with keeping the topological consistency among different graphs. Therefore, we define a node degree filter to retain the nodes with $deg(v) \geq \eta$ and $\eta$ is the node degree threshold to configure. Then, we decreasingly sort the nodes in each storage lists according to node distance to zero point (vector) in the embedding space. And we further prune the nodes in two storage lists via two more defined filters: 1) the following one is a distance filter to eliminate the nodes that have the same embedding distance to zero point from both lists, this operation helps to avoid next fast alignment among those nodes with an identical position in the embedding space; 2) the last filter will execute a fast and accurate matching to produce two sets of starting points from the storage lists for each graph, which ensures that a node that will be compressed in one graph has a ``reasonable'' counterpart for compressing in the other graph. Specifically, we repeatedly compare a node $v$ at the top of current $L_1$ with each node $u$ $\in$ top $\lambda$ of $L_2$, and the first pair with a high embedding similarity $Sim(v,u) \geqslant \omega$ are saved into $Q_1$, $Q_2$ and  removed from $L_1$, $L_2$ respectively. $\lambda$ defines the scanning range on second list for each paring operation and $\omega$ is a defined similarity threshold that determines the similar level of a matching to accept. This pairing process executes until all the nodes in either $L_1$ or $L_2$ have a match.
\begin{algorithm}[tb]
	\caption{G-CREWE($G_1$,$G_2$,$K$,$\varphi$,$\eta$,$\omega$,$\lambda$,$p$,$\gamma$,$\gamma_{1}$,$\gamma_{2}$)}  
	\begin{algorithmic}[1] 
	\vspace{0.5mm}
	    \ForAll{node $v$ $\in$ $V_1 \cup V_2$}  \Comment{Get features of all nodes}
    		 \While{$k$ $<$ $K$}
    		    \State $d_v^k$ = CreateDegreeVector($v$, $k$)
             \EndWhile 
             \State $d_v = \sum_{k=1}^{K}\gamma ^{k-1}d_v^k$
             \State Add $d_v$ to matrix $X$ \Comment{$X$ is feature matrix of all nodes}
		\EndFor
		\State $Z$ = GCNEmbed($X$, $p$) \Comment{Get structual embedding of nodes}
		\State $Z_1, Z_2$ = Split($Z$) \Comment{Split embedding for nodes in $G_1$,$G_2$}
		\State $Q_1, Q_2$ = MakeGuidingList($V_1$, $V_2$, $Z$, $\eta$, $\omega$, $\lambda$)
        \State $G'_1$, $T_1$ = MERGE($G_1, Q_1, \varphi$)
        \State $G'_2$, $T_2$ = MERGE($G_2, Q_2, \varphi$) \Comment{$T_i$ stores supernodes in $G'_i$}
        \State $Z'_1$ = CalSupEmbed($V'_1, Z_1, T_1$) \Comment{Get supernodes embedding}
        \State $Z'_2$ = CalSupEmbed($V'_2, Z_2, T_2$) 
		\State $S'$ = InferSimMat($U'_1, U'_2$) \Comment{Sim of supernodes}
		\State $S_1$ = InferSimMat($C'_1, C'_2$, $\gamma_{1}$, $\gamma_{2}$) \Comment{Sim of uncompressed nodes}
		\State $S_2$ = InferSimMat($U'_1, U'_2, S'$, $\gamma_{1}$, $\gamma_{2}$) \Comment{Sim of compressed nodes}
		\vspace*{1.2mm}
		\State =============Function MakeGuidingList=============
		\State $L_1, L_2$ = $V_1, V_2$
		\ForAll{node $v$ $\in L_1$ or $L_2$} \Comment{Filter nodes by degree}
		    \If{$deg(v) \leq \eta$}  \Comment{$\eta$: node degree threshold}
		        \State Remove $v$ from $L_1$ or $L_2$	                            
		    \EndIf
		\EndFor
		\State Sort $L_1, L_2$ by distance of node embedding to $\vec{0}$
		\State Delete nodes in $L_1, L_2$ with equal distance
		\ForAll{node $v$ $\in L_1$} \Comment{Fast pairing \& filtering}
		    \ForAll{node $u$ $\in$ top-$\lambda$ $L_2$}  
    		    \If{$Sim(v,u) \geq \omega$}  \Comment{$\omega$: similarity threshold}
		            \State Save $v$, $u$ to $Q_1$, $Q_2$
		            \State Remove $v$, $u$ from $L_1$, $L_2$
		        \EndIf
		    \EndFor
		\EndFor
		\State \Return $Q_1$, $Q_2$
		\vspace*{1.2mm}
		\State ================Function MERGE================
		\State $G'_i$ = $G_i$
		\While{compress rate $<$ $\varphi$} \Comment{$\varphi$: compression ratio}
		    \ForAll{node $v$ $\in$ $Q_i$}
		        \State New a list $L$
    		    \ForAll{node $u$ $\in$ $N(v)$}  \Comment{Neighbors of $v$ in $G'_1$}
        		    \If{$deg(u) = \delta(N(v))$}  
	 		            \State Store $u$ to $L$
		            \EndIf
		        \EndFor
		        \State Merge nodes $\in L$ as supernode $v'$ in $G'_i$ 
		        \State Store the pair of $v'$ and $L$ in $T_i$
		        \State Update $Q_i$
		    \EndFor
        \EndWhile
        \State \Return $G'_i$, $T_i$
	\end{algorithmic} 
	\label{alg:Main_alg}
\end{algorithm} 

\subsection{Node Alignments after Compression}
The final stage is to compare the similarity of the nodes in two graphs based on the Euclidean distance of embedding. When $G_1$ and $G_2$ are two original graphs, the compressed counterparts $G'_1$ and $G'_2$ are generated via MERGE in the Section \ref{sec:MERGE}. To match the nodes in $G'_1$ and $G'_2$, two similarity matrices are built for the uncompressed original nodes ($C'_1 \Leftrightarrow C'_2$) and the supernodes ($U'_1 \Leftrightarrow U'_2$), respectively. Next, the sub-nodes in those supernodes in both sides that have high similarity are further compared. If there is no node attributes involved, the way to calculate similarity score for each pair of nodes is as follows:
\begin{equation}
\begin{split}
Sim & (v, u) = exp(-Dis_1(v, u)), v \in V_1 \cup V'_1, u \in V_2 \cup V'_2
\end{split}
\label{equ:sim_cal}
\end{equation}
where $Dis_1(v, u) = \left \| Emb(v) - Emb(u)\right \|_{2}^{2}$ and the function $Emb$ can return the structural embedding for each node belonged to uncompressed nodes, supernodes or sub-nodes. However, when node attributes are available for network alignment, we use a new equation for similarity computation that is showed below:
\begin{equation}
\begin{split}
Sim(v, u) = exp(- \gamma_{1} \cdot Dis_1(v, u) - \gamma_{2} \cdot Dis_2(v, u))
\end{split}
\label{equ:sim_cal_att}
\end{equation}
Here, $Dis_2(v, u) = \left \| Fea(v) - Fea(u)\right \|_{2}^{2}$, where $Fea$ is the function to fetch attribute-based feature for each original node (compressed or uncompressed) from the attribute feature matrix $Y$ that has been established previously. Moreover, $\gamma_{1}$ and $\gamma_{2}$ are two scalar parameters that manage the weights of structural features and attribute features in similarity calculation. Note that the similarity between each pair of supernodes is still calculated with Eq. \ref{equ:sim_cal} since we do not build attributes for them in this paper. In similarities calculation, a \textit{k-d} tree is optionally applied to accelerate the process by effective comparison among nearest nodes \cite{bhatia2010survey}. Finally, we conduct \textit{Argmax} on aligning nodes in second graph to those in first graph according the similarity matrix. As an example shown in Fig.~\ref{fig:main_framwork}, supernodes $U'_1 = \{ b', f'\}$, $U'_2 = \{ 2', 6'\}$, $C'_1 = \{ d, e\}$ and $C'_2 = \{ 4, 5\}$ are formed after compression. Similarity matrix $S'$ is computed for the supernodes between $U'_1$ and $U'_2$, and matrix $S$ is for mapping the uncompressed nodes between $C'_1$ and $C'_2$. We can infer the alignments between $\{ b', f'\}$ and $\{ 2', 6'\}$ from $S'$, then compute similarity scores for the sub-nodes in these supernodes that have high similarity. One observation indicates that comparing the sub-nodes of a supernode in one graph with the sub-nodes of more top similar supernodes in another graph tends to yield better alignment result. But the trade-off between accuracy and runtime needs to be considered on the number of supernodes for comparison.

\subsection{Computational Complexity}
\label{sec:complex_ana}
There are four main stages in the proposed framework and the computational complexity is given in this section. It is different from the experiments, we assume that two graphs have a same number $n'$ of nodes. 
\begin{itemize}
\item \textbf{Node Features Extraction.} The time complexity in this process is approximately $O(n' K deg_a^2)$, which iteratively search neighbors for each node up to $K$-hop away and accumulate the neighborhood information. Usually, the average degree $deg_a$ is relative small when the graph is large.
\item \textbf{Node Embedding Learning.} We compute the time complexity for learning embedding of the nodes in two original graphs with the GCN model. \cite{wu2019comprehensive} claims that graph convolution operation computes each node's representation with involving certain amount of its neighbors, and the number of those neighbors over all nodes equals to the number of edges $|E|$. As the graph adjacency matrix often being sparse and no training required during our unsupervised learning, the main computation occurs in Eq.\ref{equ:2_layers_con} takes $O(|E|mhp)$ time. After graph compression, a certain number of supernodes are generated in each graph according to compression ratio $\varphi$. The time complexity to compute their embeddings is $O(\varphi|V_i|p)$, as it is an element-wise addition on the nodes that have been merged.
\item \textbf{Graph Compression.} The main computation in this stage happens in merge operations. One merge action first identifies the adjacent nodes $L$ of a start point $v_{start}$ that have minimum degree among $N(v_{start})$. Then, nodes in $L$ are removed from the graph and new connections are created between $N(L)$ and a supernode. Therefore, the complexity of this operation is around $O(deg_a |L|_{a})$. The average size of minimum degree neighbors $|L|_{a}$ is smaller than average degree $deg_a$. The total number of merging is $\frac{\varphi|V_i|}{|L|_{a}}$, which makes the final complexity equal to $O(\varphi|V_i| deg_a)$.
\item \textbf{Node Alignments.} The \textit{k-d} tree could be used to find the top alignment(s) in $G'_2$ for either each uncompressed node or supernode in $G'_1$ in average time complexity $O(|C'_1| log|C'_1|)$ and $O(|U'_1| log|U'_1|)$, respectively. As we known, $|C'_1| + |U'_1| = |V_1|(1-\varphi)$. 
\end{itemize}
$O(max\{n' K deg_a^2, |E|mhp, \varphi|V_i|p, \varphi|V_i| deg_a, |C'_1|log|C'_1|+|U'_1|log|U'_1|\})$ is the total complexity of the framework. In the experiments, $K, m, h$ and $p$ are often configured to small values.

\section{Experimental Results}
\label{sec:experiment}
In this section, we evaluate the performance of the proposed algorithm with the baselines in two main aspects: alignment accuracy and CPU runtime. In addition, the effect of different hyper-parameters on the performance of the G-CREWE is further investigated.

\subsection{Experimental Setup}
\textbf{Machine.} Our algorithm is implemented in Python 3.7 and the GCN model is CPU consuming for embedding calculation. All the experiments are executed on a machine with Intel(R) Xeon(R) Gold 6132 CPU @ 2.60GHz and Memory 500 GB.
\\ 
\textbf{Dataset.} Three different types of dataset are used for experimental evaluation. The statistics of the networks are summarized in Table \ref{datasets}. Following the network alignment literature \cite{zhang2017ineat,heimann2018hashalign,heimann2018regal}, given a network $G_1$ with adjacency matrix $A_1$, we create a network alignment problem with known ground truth: a noisy permutation $G_2$ with adjacency matrix $A_2 = PA_1P^T$, where $P$ is a randomly generated permutation matrix. Different level of structural noise is then added to $A_2$ by removing each edge $e \in E_1$ with a defined probability. For testing the performance of algorithms on more disturbed networks, we accumulate the edge noise at different levels on the permuted network step by step. Noticeably, the nodes with zero degree in a noisy network are discarded during edge elimination, which might produce the networks with distinct size of nodes or edges in alignment. As for the experiments with attributes, we produce synthetic categorical attributes for each node in both graphs. Every node has a same number of attributes and the noise is added via replacing each original attribute value with a different category uniformly at random with a defined probability.
\\ 
\textbf{Baselines.} We mainly evaluate the alignment performance of the proposed solution and the other baselines that had not considered both node embedding and graph compression. We consider two classic baselines, the message-passing optimization algorithm \textbf{\textit{NetAlign}}~\cite{bayati2013message} and the spectral method \textbf{\textit{IsoRank}}~\cite{singh2008global}. \textbf{\textit{REGAL}} \cite{heimann2018regal} is one recent network alignment method, which is comparable to our approach. REGAL applies node representation to perform network alignment and we use the default setting for it. In \cite{heimann2018regal}, REGAL is reported to be much more superior against other algorithms, including FINAL \cite{zhang2016final}, in both the efficiency and effectiveness measures. Therefore, in this paper, we assume the transitivity of this performance, and therefore REGAL is chosen as one of the key benchmark algorithms. In addition, two variants of the proposed framework \textbf{\textit{G-CREWE-xNetMF}} and \textbf{\textit{G-CREWE-Shrink}} are implemented for comparative analysis. The former replaces the GCN embedding approach in G-CREWE with xNetMF \cite{heimann2018regal}, which is an efficient matrix factorization-based method that extends low-rank approximation for multiple input networks. For xNetMF method, we consider a maximum hop distance $K = 2$ and the number of landmarks is $10 \times log(|V|)$. The latter one use a novel graph compression algorithm called Shrink \cite{sadri2017shrink} as the substitute of MERGE but keeping the section of guiding-list generation. Shrink can effectively coarsen the graph with a defined compression ratio while preserving the path distance between the nodes. To use it in our framework, we compress each pair of nodes in each merge operation with its node selection criteria. The edge weights is not considered in this problem and the weights of new connections between a super-node and its neighbors is set to one. The initial threshold $\theta$ for node selection of compression is defined to one and dynamically updated by condition $(N(u) + N(uv)) \times (N(v) +N(uv)) < \theta$, where $u, v$ are two adjacent nodes in a network. 
\\ 
\textbf{Evaluation Metrics.} Two metrics are used to compare our method to the baselines: (1) alignment accuracy [($\#$ correct alignments) / ($\#$ total alignments)] and runtime; (2) top-$\alpha$ alignment accuracy [($\#$ correct alignments in top-$\alpha$) / ($\#$ total alignments)]. And runtime is recorded simultaneously. We executed five independent trials on each dataset in different settings and compared mean of the results.

\begin{table}[tb]
\caption{Description of the datasets for experiments}
\begin{center}
\begin{tabular}{llll}
\toprule
\textbf{Name} & \textbf{\# Nodes} & \textbf{\# Edges} & \textbf{Description} \\
\hline 
Enron \cite{leskovec2009community} & $35,235$ & $183,082$ & Communication network \\
Brightkite \cite{cho2011friendship} & $57,458$ & $214,078$ & Social network \\
DBLP \cite{yang2015defining} & $111,812$ & $351,577$ & Co-authorship network \\
\bottomrule 
\end{tabular}
\label{datasets}
\end{center}
\end{table}

\subsection{Alignment Performance Analysis}
\label{sec:alignment_analysis}
In this section, we study the alignment accuracy and runtime of G-CREWE and the baselines on three different types of datasets with different level of edge noise and node attribute noise. Noise zero indicates that the second network is the permutation of first one without adding noise. The larger edge noise is gradually accumulated on the network with previous noise, which will bring more challenges to our alignment algorithm with compression. G-CREWE uses structural features of nodes to build network embedding and can also combine it with node attributes for final network alignment. Main hyper-parameters in this experiment are as follows: maximum hop distance $K = 2$; hop discount factor $\gamma = 0.01$; compression ratio $\varphi = 0.2$; GCN hidden matrix dimension $m = log_{2}\Delta(G)$ and $h = 16$; GCN embedding dimension $p = log_{2}\Delta(G) \times 2$; $\gamma_1 = \gamma_2 = 1$. In addition, the parameters in the stage of guiding-list generation are node degree threshold $\eta = 15$, number of top nodes for fast pairing $\lambda = 100$ and node similarity threshold $\omega = 0.98$. \\
\begin{figure*}[tb]
    \centering
    \subfloat[DBLP(111,812 nodes) \label{subfig:DBLP_acc_run_time}]{%
    \includegraphics[width=0.32\textwidth]{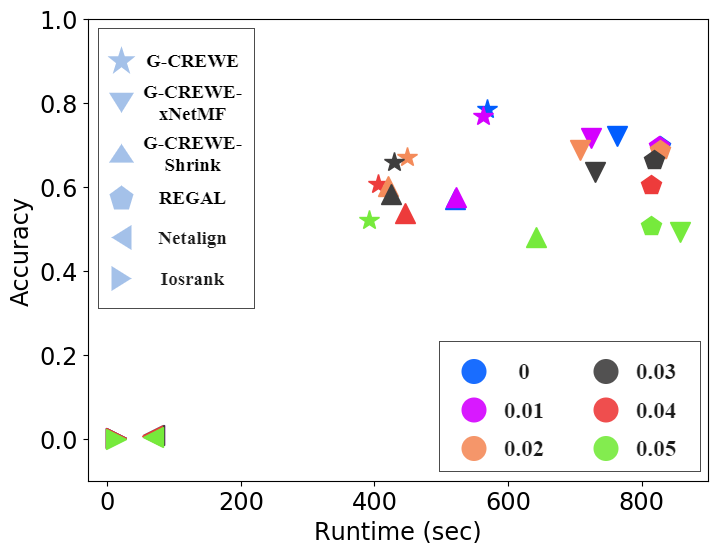}
    }
    \subfloat[Brightkite(57,458 nodes) \label{subfig:Brightkite_acc_run_time}]{%
    \includegraphics[width=0.32\textwidth]{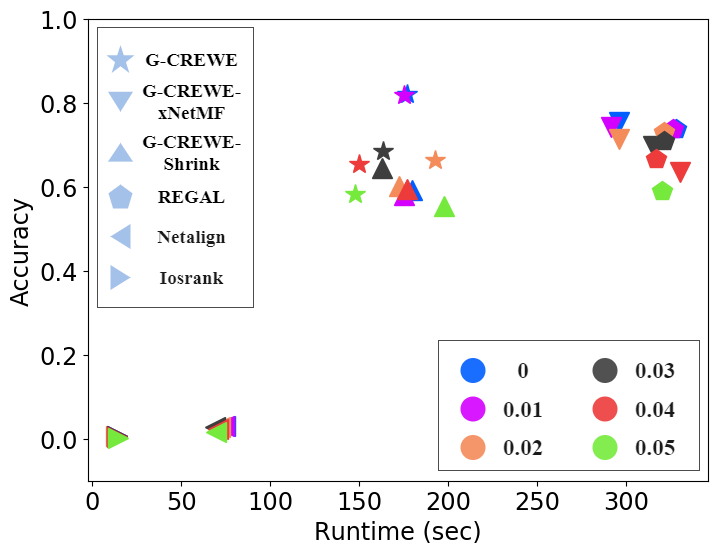}
    }
    \subfloat[Enron(35,235 nodes) \label{subfig:Enron_acc_runtime}]{%
    \includegraphics[width=0.32\textwidth]{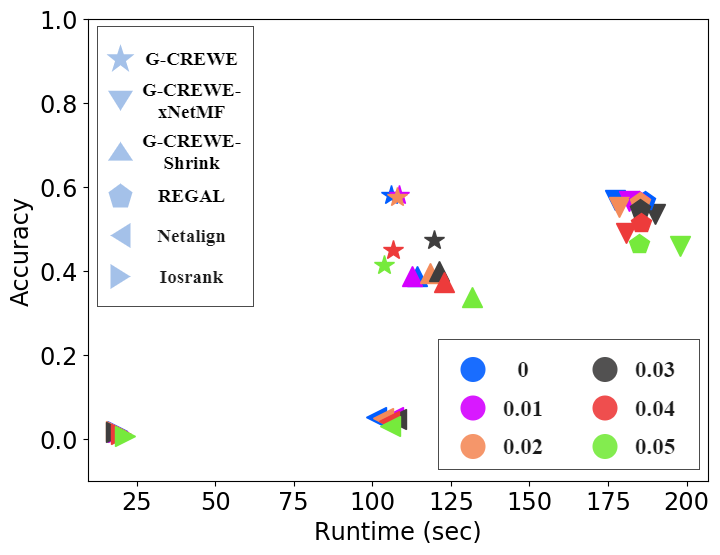}
    }
    \caption{Runtime vs accuracy for alignment with different edge noise on DBLP, Brightkite and Enron datasets. Across noise levels, G-CREWE offers the best combination of accuracy and runtime.  Moreover, G-CREWE is flexible enough to work with different node embedding and compression methods, but the combination of GCN and proposed MERGE is the fastest and still very accurate.}
    \label{fig:acc_run_time}
\end{figure*}
\textbf{Effects of Edge Noise.} In this evaluation, the representation-based approaches (G-CREWE, REGAL and two proposed variants) apply a \textit{k-d} tree to search the top alignments in second network, which could avoid global comparison on nodes for saving memory and query time \cite{heimann2018regal}. Fig.~\ref{fig:acc_run_time} compares all the methods on their alignment \emph{efficiency}: not only how accurate are the alignments they find, but also how quickly they find them.  In our plots of runtime versus accuracy, the first finding is that the methods IsoRank and NetAlign are significantly less accurate, occupying the lower portions of the plot. One reason for this is that they cannot reserve or capture node structural information precisely and their performance seems being worsen when the networks become relative large. Comparing G-CREWE to REGAL, the most competitive baseline, we see that the two methods are similar in accuracy. However, G-CREWE is significantly faster than REGAL across datasets and at all levels of noise, as is reflected by its location closer to the upper left of the plot (the most desirable region, reflecting low runtime and high accuracy).  This indicates a higher alignment efficiency of G-CREWE in that it can maintain competitive accuracy while offering large (approximately two times) speedups as a result of the compression. 

\begin{figure*}[tb]
    \centering
    \subfloat[3 synthetic binary attributes \label{subfig:acc_dif_noi_level_3_attrs}]{%
    \includegraphics[width=0.325\textwidth]{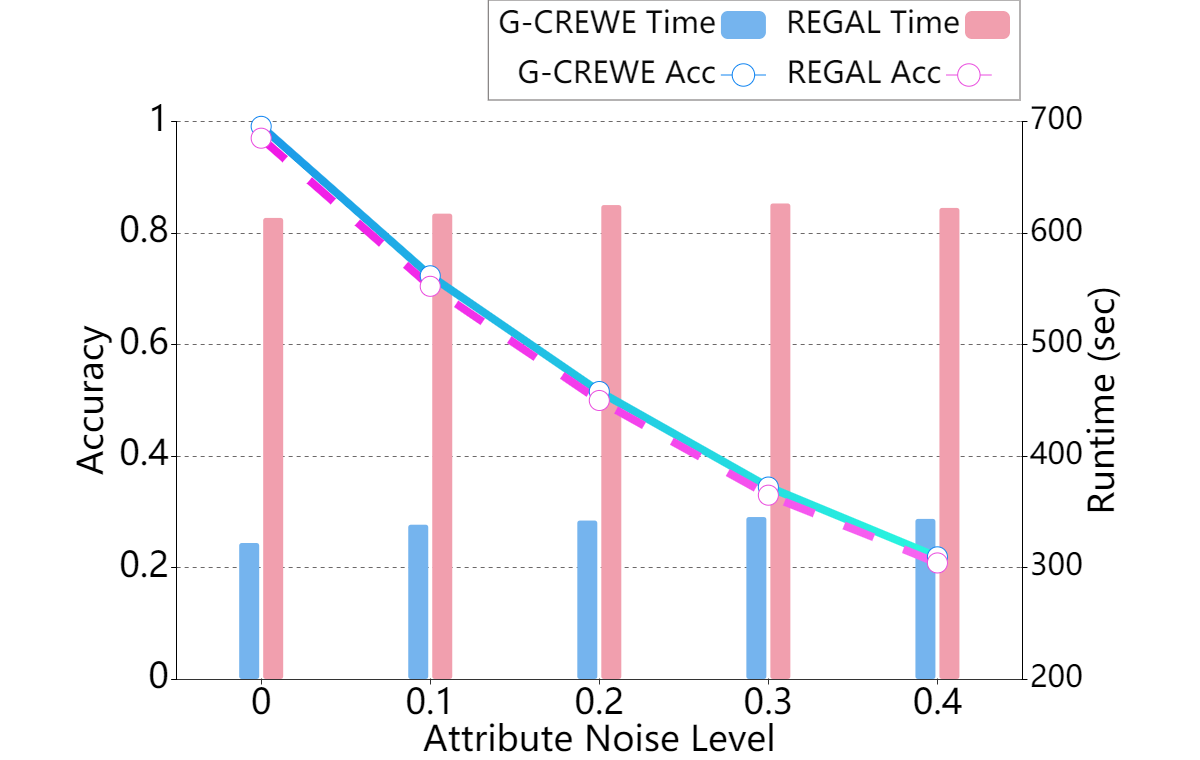}
    }
    \subfloat[5 synthetic binary attributes \label{subfig:acc_dif_noi_level_5_attrs}]{%
    \includegraphics[width=0.325\textwidth]{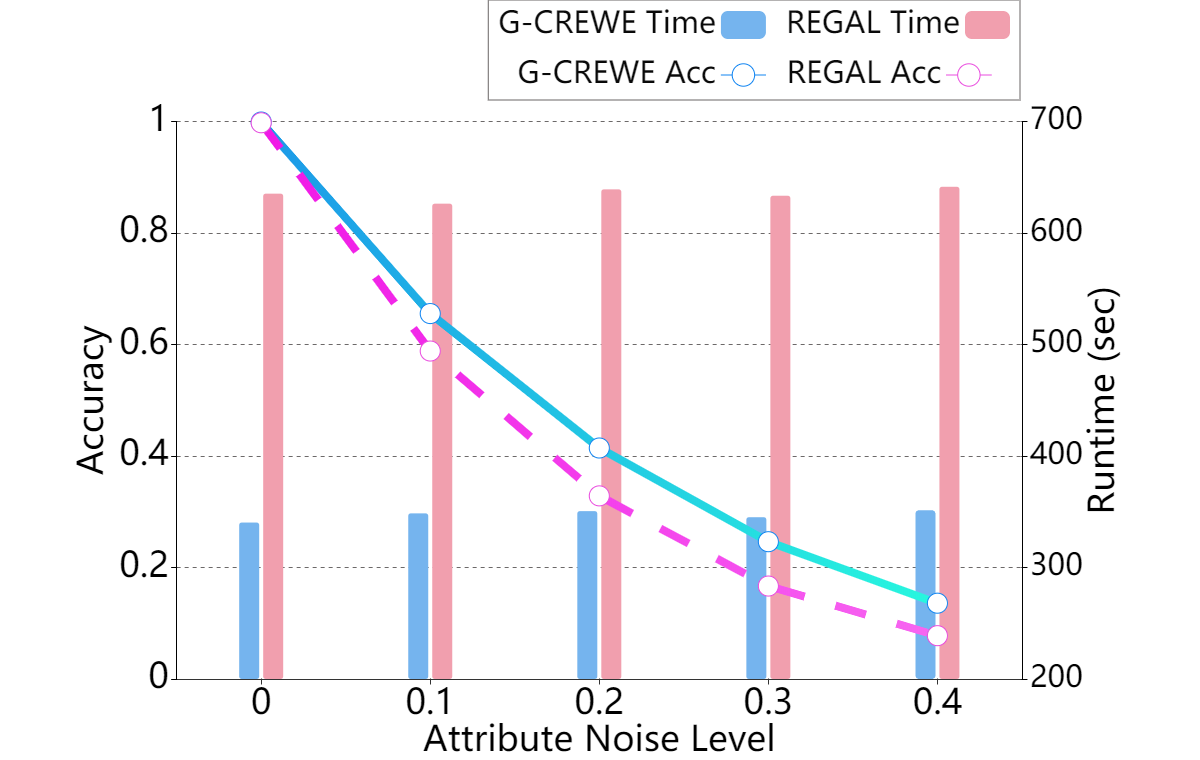}
    }
    \subfloat[7 synthetic binary attributes \label{subfig:acc_dif_noi_level_7_attrs}]{%
    \includegraphics[width=0.325\textwidth]{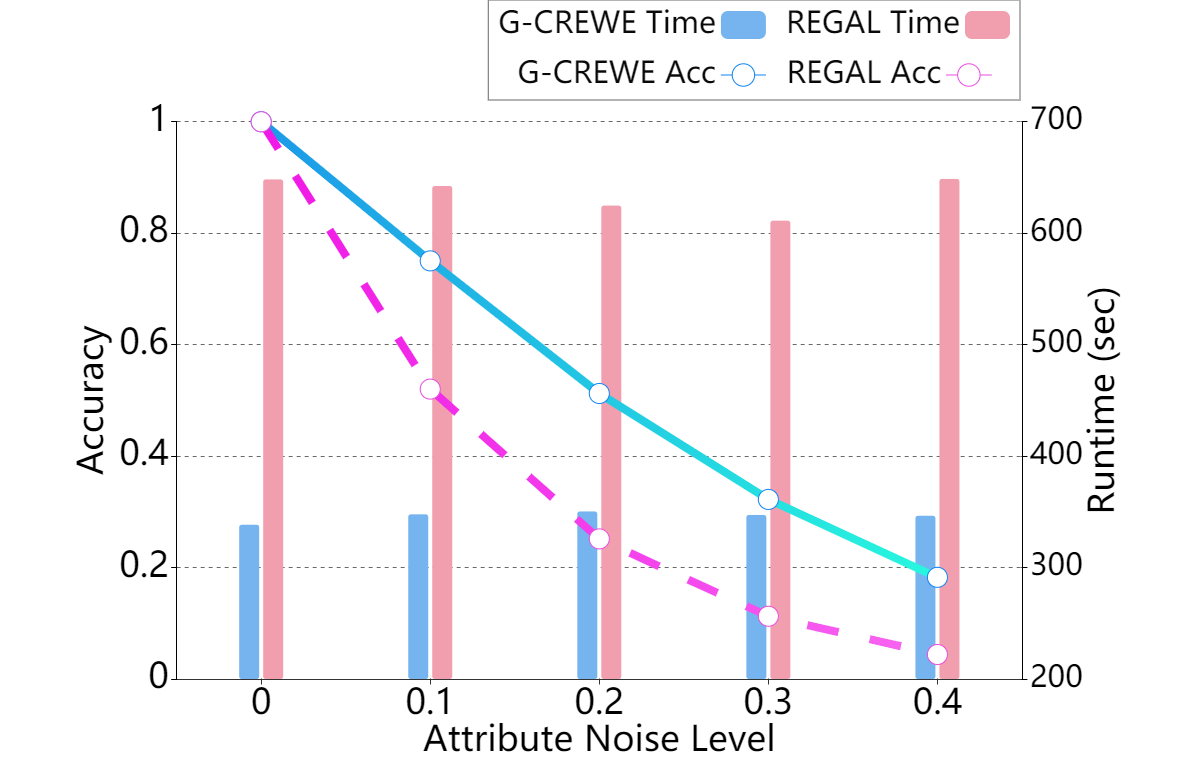}
    }
    \caption{Average alignment accurate and runtime on Brightkite network with different levels of node attribute noise. In this setting, the edge noise is 0.01 for all the tests with 3, 5 or 7 binary node attributes respectively. G-CREWE is faster and shows stronger reliability in alignment performance with an increase of attribute number and noise.}
    \label{fig:dif_attrs_noi_level_acc}
\end{figure*}

We can see that G-CREWE and the other embedding-based methods experience a gradual decrease in alignment accuracy as edge noise increases. It is noticeable that the alignment accuracy remains relatively high for G-CREWE on both Brightkite and DBLP when the noise level is between $0$ to $0.02$, but it drops to around 0.6 when more noise is added. On contrast, REGAL has moderate alignment accuracy in this experiment. NetAlign and IsoRank are two stable alignment methods but the accuracy is significantly below the embedding-based methods across noise levels. The runtime of G-CREWE keeps almost twice faster than that of REGAL on three datasets. For instance, it consumes around $400$ to $500$ seconds for each alignment on DBLP with $111,812$ nodes and the baseline REGAL usually needs about 800 seconds to finish one test. 
\\
\textbf{Effects of Attribute Noise.} Node attributes could be integrated with structural feature for network alignment. We compare G-CREWE with REGAL which is the well-established baseline that consider both types of features. A difference from the test of edge noise, we discard the use of kd-tree for conducting a global comparison between nodes of two networks, which tends to yield higher accuracy. Fig. \ref{fig:dif_attrs_noi_level_acc} shows the average alignment accurate and runtime on Brightkite network with different levels of node attribute noise. In this setting, the edge noise is 0.01 for all the trials with 3, 5 or 7 binary node attributes respectively. G-CREWE is faster and shows stronger reliability in alignment performance with an increase of either node attribute number or noise. We can also see that, the alignment accuracy becomes slightly higher when more node attributes are considered without any attribute noise, but the figure of G-CREWE drops gradually from approximately 0.97 to 0.2 while the noise is increasing and the baseline experiences an even worse decrease. Compared to Fig. \ref{subfig:Brightkite_acc_run_time} that only uses structural information, the best accuracy among all the algorithms is just 0.85 while the edge noise is same. Overall, the performance of G-CREWE is more stable than the baseline when the attribute number is raised with same degrees of noise.
\\
\textbf{Variant Methods.} G-CREWE-xNetMF and G-CREWE-Shrink are two variants created under the proposed framework, which uses either a different embedding method or compression method. By studying these two competitive methods with G-CREWE, we can gain two insights: first, we can see if the framework works well with different kinds of node embeddings or compression approaches, and second, we obtain a breakdown that the performance benefits yielded by each separate stage of the proposed framework. Using GCN in G-CREWE is competitive to xNetMF in alignment accuracy, but the speed of the graph convolution operations consistently bring the former an advantage in computational time compared to the element-wise construction of the factor matrices in xNetMF. Moreover, the combination of GCN and Shrink (G-CREWE-Shrink) can also boost the overall runtime compared with non-compression method REGAL, but it experiences a drop in alignment accuracy. In contrast, using MERGE with GCN could ensure higher accuracy.

\begin{figure*}[tb]
    \centering
    \subfloat[Permuted network without edge noise \label{subfig:Brightkite_ratio_acc_run_time_0}]{%
    \includegraphics[width=0.33\textwidth]{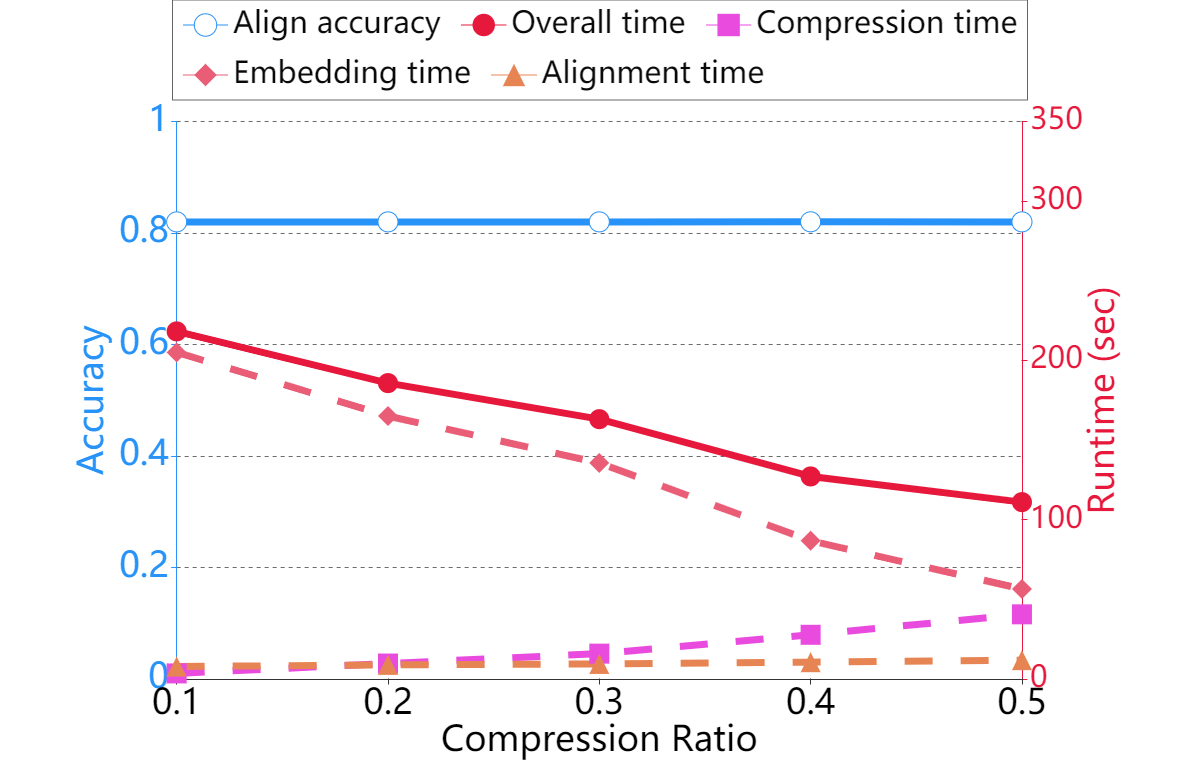}
    } 
    \subfloat[Permuted network with 0.01 edge noise \label{subfig:Brightkite_ratio_acc_run_time_01}]{%
    \includegraphics[width=0.33\textwidth]{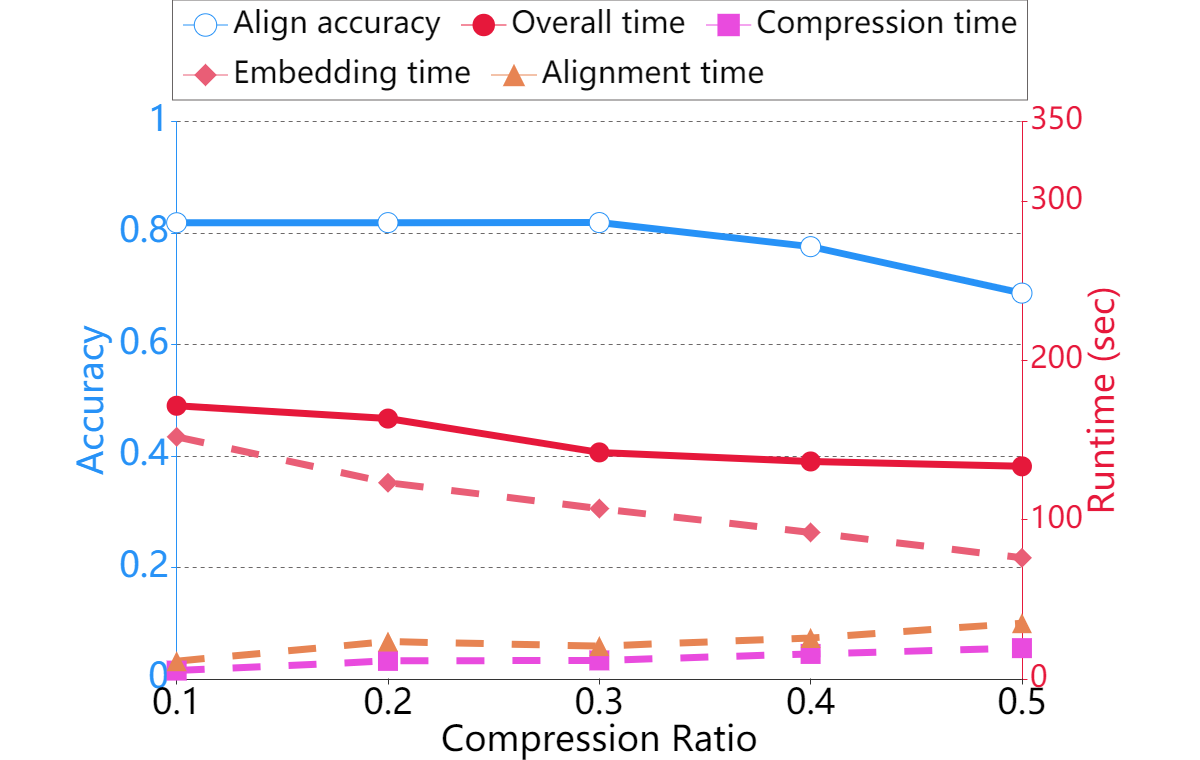}
    }  
    \subfloat[Top-$\alpha$ scores \label{subfig:Brightkite_top_k_acc_bar}]{%
    \includegraphics[width=0.32\textwidth]{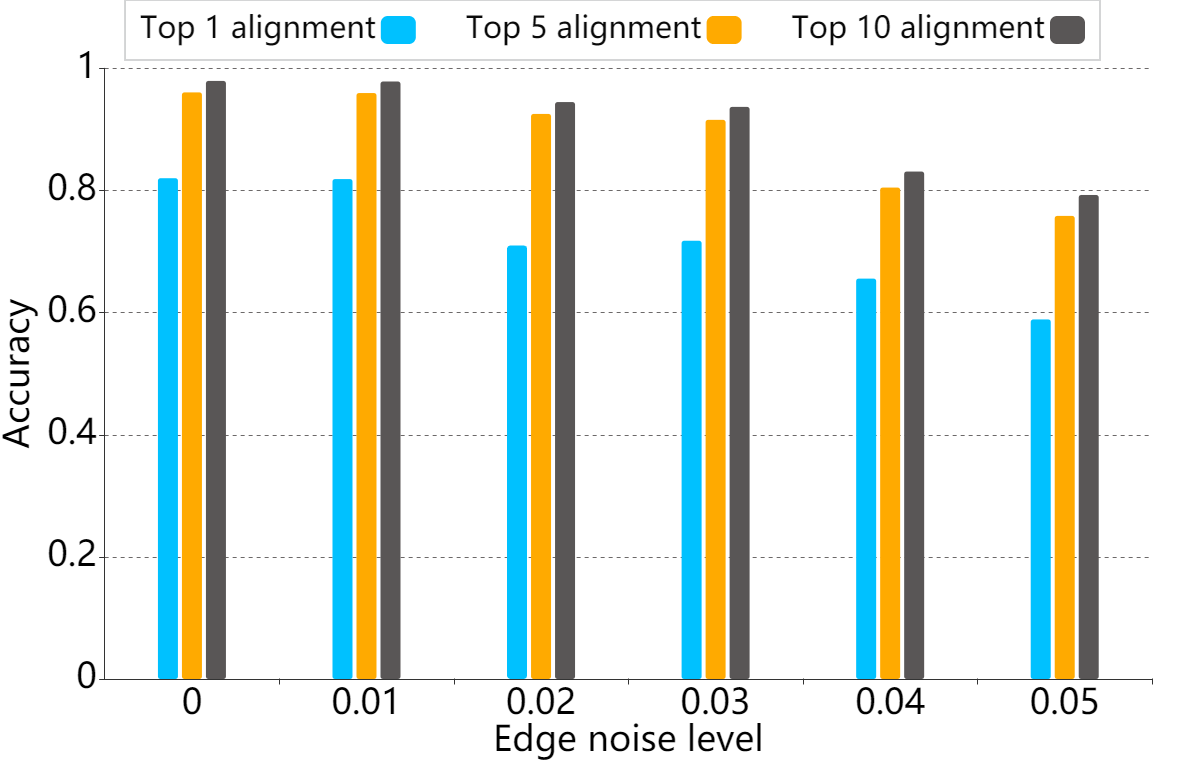}
    }
    \caption{Average alignment accuracy and runtime of different stages of G-CREWE with varied compression ratio, along with top-$\alpha$ scores, on Brightkite dataset. (a) and (b) show that moderate compression provides the best runtime trade-off and high accuracy without or with minor edge noise. (c) G-CREWE can match even more nodes when the alignment criterion is relaxed to top-$\alpha$.}
    \label{fig:Brightkite_ratio_acc_run_time_}
\end{figure*}

\balance
\subsection{Sensitivity Analysis}
\begin{figure}[tb]
    \centering
    \subfloat[Accuracy of using different number of layers in GCN \label{subfig:layer_acc_bar}]{%
    \includegraphics[width=0.7\linewidth]{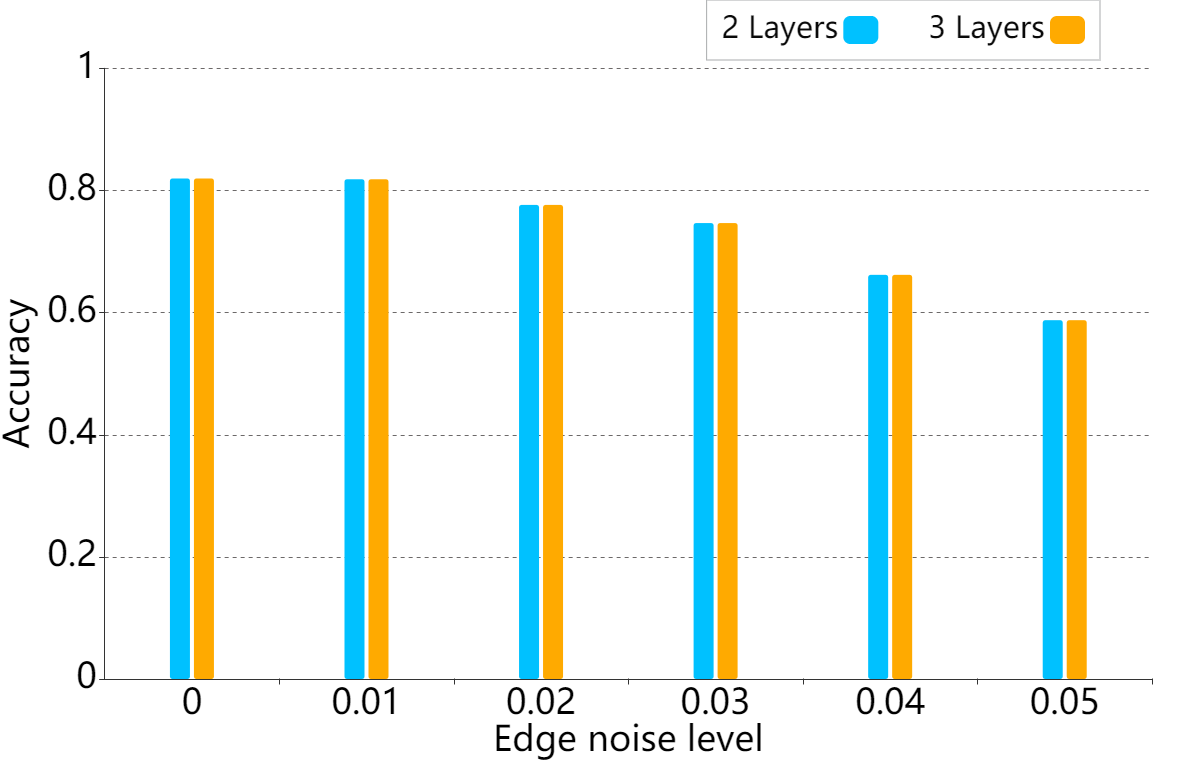}
    } \\
    \subfloat[Runtime of using different number of layers in GCN \label{subfig:layer_runtime_bar}]{%
    \includegraphics[width=0.7\linewidth]{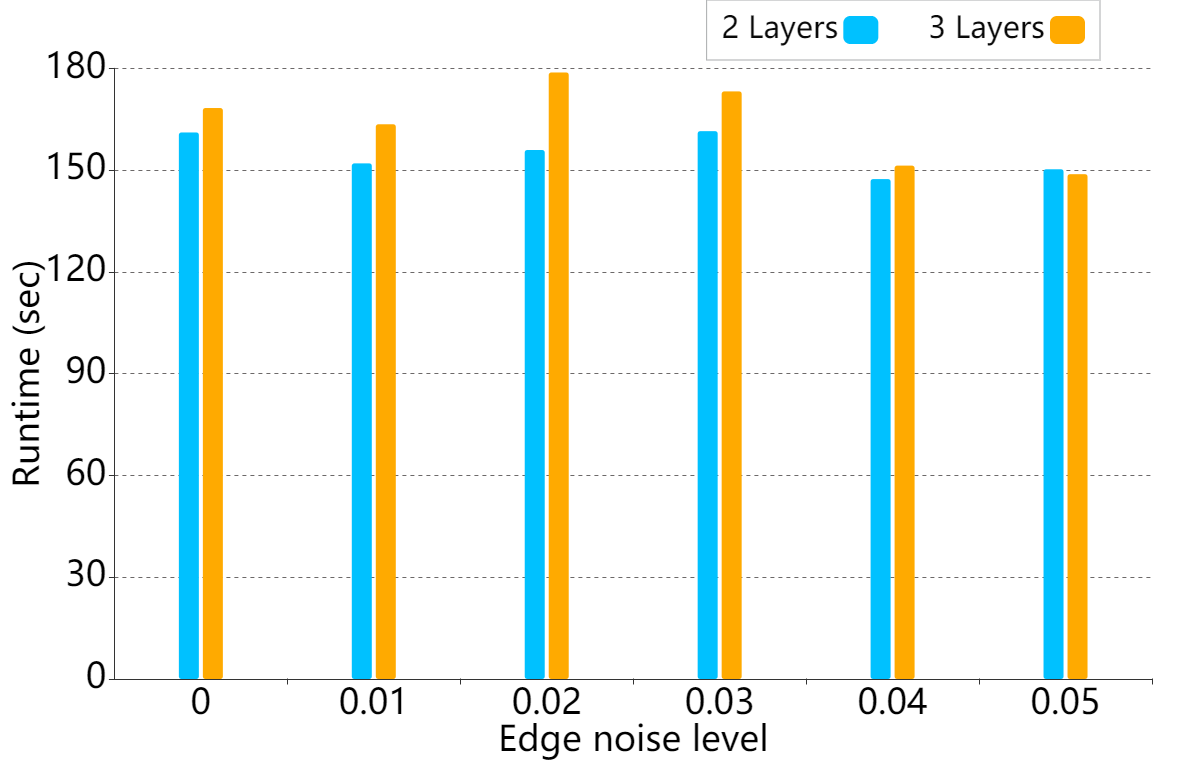}
    }  
    \caption{Accuracy and runtime of using different number of layers in GCN for node alignment on Brightkite.}
    \label{fig:layer_accu_runtime_}
\end{figure}

In this section, we examine the affect of some main hyper-parameters on the performance of G-CREWE in several experiments. This could provide us more insights about the advantages of the proposed algorithm and the possible limitations in different settings.
\\
\textbf{Compress Ratio Trade-off.} We first evaluate the impact of compression ratio on alignment accuracy and runtime with dataset Brightkite. Fig.~\ref{fig:Brightkite_ratio_acc_run_time_}a shows the alignment accuracy is stable when we increase the compression ratio from 0.1 to 0.5 on the permuted network but without edge noise. Meanwhile, the overall runtime reduces correspondingly from more than 200 seconds to approximately 110 seconds. The main reason is that the runtime of embedding learning decreases dramatically when we intensively compress a network into smaller one. However, the figure becomes different when edge noise is added to the permutated network. As we can see from Fig.~\ref{fig:Brightkite_ratio_acc_run_time_}b, the accuracy maintains high with a compression ratio $\varphi \leq 0.3$, but it drops obviously along with the growth of overall runtime while we increase the number of compressed nodes. That is because increasing the amount of node and edge elimination in noisy networks can prevent the method on gaining more high-quality starting points in building the guiding-list. Subsequently, a large number of sub-nodes may be merged into one supernode for achieving high compression ratio. In addition, the structural consistency is harder to maintained between both compressed networks.  Therefore, matching a big size of sub-nodes between correspondingly similar supernodes could slow down the overall computation time. And intensive compression on noisy networks could further deteriorate the alignment accuracy due to the compressed nodes match locally between their highly similar supernodes and lack a full view on far ones. As a result, moderate compression provides the best runtime trade-off and high accuracy. 
\\
\textbf{Top $\alpha$ Accuracy.} Going beyond ``hard'' node alignments, we also study the top-$\alpha$ scores of ``soft'' alignments. The setting of parameters in this evaluation is same with Section \ref{sec:alignment_analysis}. Fig.~\ref{fig:Brightkite_ratio_acc_run_time_}c demonstrates the results for $top-1$, $top-5$, and $top-10$ alignments on Brightkite with different noise levels. A higher $\alpha$ results in even more successful alignments, which is useful in some applications such as making recommendations to more than one user in one network when giving the information in another network.
\\
\textbf{Number of Layers.} On the assessment of using different number of layers in GCN model for embedding learning, Fig.~\ref{fig:layer_accu_runtime_} demonstrates that layer 2 and layer 3 could contribute almost equally on the aspect of alignment accuracy, but the runtime of layer 3 is slightly larger than using smaller number of layers.
\\
\textbf{Other Parameters.} In extracting structural features of a node, $\gamma$ controls the impact of its neighbors according to distance. Nearer neighbors have more influence if $\gamma < 0$ and vice versa. Some new parameters are also introduced in this paper, such as node degree threshold $\eta$, number of top nodes for fast pairing $\lambda$ and node similarity threshold $\omega$ in the stage of guiding-list generation. Specifically speaking, $\eta$ can control the level of ignoring nodes with low degree that are not ideal starting points for producing supernodes in compression, but less number of starting nodes could be found if $\eta$ is high. In searching starting nodes, $\lambda$ is used to define the number of top ranked nodes in the second node list that will be compared with each node $v$ in the first node list, and $\omega$ is the similarity threshold to accept a matching. A larger $\lambda$ could let node $v$ to compare with the nodes that are further away in embedding space but consume more computational time, and a low $\omega$ may worsen the quality of finding pairs of similar starting points from both networks. 

\section{Related Work}
\label{sec:relate_work}
Many researches have shown their extensive work on solving network alignment. IsoRank \cite{singh2008global} and NetAlign \cite{bayati2013message} are two classical alignment methods, the former infers pairwise node similarity in multiple PPI networks and the latter designs a new message passing algorithm for the sparse network alignment. More recently, HubAlig \cite{hashemifar2014hubalign} uses both network topology and sequence homology information to align PPI networks and ModuleAlign \cite{hashemifar2016modulealign} calculates cluster-based homology score to improve the alignment performance immediately after. Zhang and Tong developed a series of algorithms called FINAL to match attributed networks \cite{zhang2016final}. In 2018, REGAL applies xNetMF to generate effective node representations for node alignments \cite{heimann2018regal}. Moana, a multilevel approach, was developed by coarsening the input networks into multiple levels and finding the alignment with the patterns at different phases \cite{zhang2019multilevel}.

A large amount of work have been conducted on the compression of Web graph and social networks. Chierichett et al. utilize link reciprocity in social networks for compression with consideration of node ordering \cite{chierichetti2009compressing}. Similarly, Boldi et al. also reorder the nodes in graphs for faster compression with WebGraph framework \cite{boldi2011layered}. Fan et al. proposed query preserving graph compression to reduce the size of graphs with ensuring the correctness in reachability queries and pattern matching \cite{fan2012query}. Toivonen et al. proposed an generalized weighted graph compression problem and solved it with a range of algorithms \cite{toivonen2011compression}. Shrink is a distance preserving graph compression algorithm that speeds up the searching of shortest path and save more space \cite{sadri2017shrink}. 

Node embedding learning aims at finding representative features for nodes in graph. DeepWalk \cite{perozzi2014deepwalk} generates continuous feature representations for the nodes in networks through turning a graph structure into linear sequences via truncated random walks. LINE \cite{tang2015line} preserves both the local and global network structures with the optimization on a designed objective function. Various generalizations such as node2vec \cite{grover2016node2vec} or struc2vec \cite{ribeiro2017struc2vec} are designed to capture the neighborhood structure of each node in different ways. Wang et al. proposed a method called SDNE to save the highly non-linear network structure via first-order and second-order proximity \cite{wang2016structural}. Qiu et al. conducted a theoretical analysis of several embedding approaches and proposed NetMF to explicitly factorize the closed-form matrices \cite{qiu2018network}. In addition, recent researches show that GCNs are capable of extracting node features for arbitrarily structured networks and perform efficiently in different learning missions \cite{kipf2016semi,chen2018harp}. 

\section{Conclusion}
\label{sec:conclusion}
The proposed framework G-CREWE fuses the processes of graph compression and network alignment together by the leverage of node embedding. The framework allows them assist each other and effectively boost the overall computational speed. Inspired by some applications of using node representation for network alignment or matching, we employ a model of GCN to complete node embedding learning and the embedding is used to supervise the compression operations in order to shorten the alignment time on coarse networks. In addition, the proposed framework also allows the usage of node attributes for alignment purpose via combining structural embedding  with encoded attribute features during matching stage. There is a trade-off between the compression degree, runtime and alignment accuracy, especially when more edge noise is added to networks. The experiments have provided several new insights relative to this concern. With appropriate compression on the network in terms of coarsening degree and node selection, we can achieve a relatively fast alignment with reserving high accuracy.

\begin{acks}
This research is supported by Australian Research Council Discovery grant DP190101485.
\end{acks}
\bibliographystyle{ACM-Reference-Format}
\bibliography{main}


\end{document}